\documentclass[a4paper,UKenglish]{lipics-v2016}
\usepackage{microtype}%if unwanted, comment out or use option "draft"
\usepackage[vlined,algo2e,ruled]{algorithm2e}

\newcommand{\rank}{\ensuremath{\mathsf{rank}}}
\newcommand{\select}{\ensuremath{\mathsf{select}}}
\newcommand{\match}{\ensuremath{\mathsf{match}}}
\newcommand{\parent}{\ensuremath{\mathsf{parent}}}
\newcommand{\first}{\ensuremath{\mathsf{first}}}
\newcommand{\Next}{\ensuremath{\mathsf{next}}}
\newcommand{\mate}{\ensuremath{\mathsf{mate}}}
\newcommand{\vertex}{\ensuremath{\mathsf{vertex}}}
\newcommand{\counting}{\ensuremath{\mathsf{counting}}}
\newcommand{\listing}{\ensuremath{\mathsf{listing}}}
\newcommand{\face}{\ensuremath{\mathsf{face}}}

\newcommand{\asgn}{\mathrel{=}}
\newcommand{\parAlgo}{{\sc par-spe}}

\title{Parallel Construction of Compact Planar Embeddings\footnote{The second
    and fifth authors received travel funding from EU grant H2020-MSCA-RISE-2015
    BIRDS GA No. 690941. The second author received funding from Conicyt
    Fondecyt 3170534.
    The third and fifth authors received funding Basal
    Funds FB0001, Conicyt, Chile. The third author received funding from Academy
    of Finland grant 268324. Early parts of this work were done while the third
    author was at the University of Helsinki and while the third and fifth
    authors were visiting the University of A Coruña.}}
\titlerunning{Parallel Construction of Compact Planar Embeddings} %optional, in case that the title is too long; the running title should fit into the top page column

\author[1]{Leo Ferres}
\author[2]{Jos\'e Fuentes-Sep\'ulveda}
\author[3]{Travis Gagie}
\author[4]{Meng He}
\author[2]{Gonzalo Navarro}
\affil[1]{Faculty of Engineering, Universidad del Desarrollo, Chile\\
  \texttt{lferres@udd.cl}}
\affil[2]{Department of Computer Science, University of Chile, Chile\\
  \texttt{jfuentess@dcc.uchile.cl, gnavarro@dcc.uchile.cl}}
\affil[3]{School of Computer Science and Telecommunications, Diego Portales University, Chile\\
  \texttt{travis.gagie@mail.udp.cl}}
\affil[3]{Faculty of Computer Science, Dalhousie University, Canada\\
  \texttt{mhe@cs.dal.ca}}
\authorrunning{J.\,Q. Open and J.\,R. Access} %mandatory. First: Use abbreviated first/middle names. Second (only in severe cases): Use first author plus 'et. al.'

\Copyright{Leo Ferres, Jos\'e Fuentes-Sep\'ulveda, Travis Gagie, Meng He and Gonzalo Navarro}%mandatory, please use full first names. LIPIcs license is "CC-BY";  http://creativecommons.org/licenses/by/3.0/

% mandatory: Please choose ACM 1998 classifications from http://www.acm.org/about/class/ccs98-html . E.g., cite as "F.1.1 Models of Computation". 
% mandatory: Please provide 1-5 keywords
\EventEditors{John Q. Open and Joan R. Acces}
\EventNoEds{2}
\EventLongTitle{42nd Conference on Very Important Topics (CVIT 2016)}
\EventShortTitle{CVIT 2016}
\EventAcronym{CVIT}
\EventYear{2016}
\EventDate{December 24--27, 2016}
\EventLocation{Little Whinging, United Kingdom}
\EventLogo{}
\SeriesVolume{42}
\ArticleNo{23}
\begin{document}

\maketitle

\begin{abstract}
The sheer sizes of modern datasets are forcing data-structure designers to
consider seriously both parallel construction and compactness. To achieve those
goals we need to design a parallel algorithm with good scalability and with low
memory consumption. An algorithm with good scalability improves its performance
when the number of available cores increases, and an algorithm with low memory
consumption uses memory proportional to the space used by the dataset in
uncompact form. In this work, we discuss the engineering of a parallel algorithm with linear work
and logarithmic span for the construction of the compact representation of
planar embeddings. We also provide an experimental study of our implementation and prove
experimentally that it has good scalability and low memory
consumption. Additionally, we describe and test experimentally queries supported
by the compact representation.

 \end{abstract}

\section{Introduction}
\label{sec:introduction}
Planar embeddings are present in several applications that need an
underlying representation of topological information, such as in the
mesh representation in finite-element simulations, road networks and
Geographical Information Systems (GIS) in general. Because of their
very nature, the size of such planar embeddings is large and dynamic,
meaning it is constantly growing. For example, the underlying planar
embedding to represent {\em OpenStreetMap}, has more than 3 billion
nodes\footnote{In OpenStreetMap, a node is defined as a specific point
  on the earth's surface defined by its latitude and
  longitude. Updated statistics are available at
  \url{http://www.openstreetmap.org/stats/data\_stats.html} (last
  access: April 06, 2017)}. Manipulating those large planar
embeddings, storing and updating them efficiently is of practical
importance. Having a data structure with a small memory footprint to
represent this class of graphs will help us manipulate huge graphs in
small devices with little (on-board, fast) memory.

Since the 1990s, compact data structures have become a viable
alternative to represent large data in a small space, storing data in
space close to its information-theoretic lower bound while still
supporting queries efficiently.  Their usefulness has been practically
demonstrated in several libraries, such as, {\em LibCDS}
\cite{libcds}, {\em SDSL} \cite{sdsl}, {\em Sux} \cite{sux}, {\em
  Dynamic} \cite{dynamicLib} and {\em Succinct} \cite{ot}. However,
there has been no attempt to implement and test the practical
efficiency of compact data structures for planar graphs with planar
embeddings.

The construction stage of compact data structures is of particular
interest, since fast construction algorithm can simulate real-time
updates, effectively making it behave as a dynamic version of it. In
this work, we take advantage of multicore architectures to provide a
fast implementation of a algorithm to construct compact data structures for planar
embeddings.

Recently, Ferres et al.~\cite{WADSpaper} extended the compact representation of planar embeddings of Tur\'an \cite{Turan1984}. The extension supports 
navigation queries, without greatly increasing the complexity introduced in the original. In the same work, Ferres et al. also introduced a 
work-optimal parallel algorithm with logarithmic depth for the extended compact representation. In our work, we provide the algorithm engineering of 
the parallel algorithm of Ferres et al. We discuss the implementation of each used parallel algorithm, such us, Euler Tour and spanning tree, and 
discuss some practical trade-offs. We provide a set of experiments to prove the scalabitily and good space-usage of our implementation, using a small 
portion of the original input. Finally, we also provide implementations of useful queries that also behave efficiently.

The layout of the rest of the paper is as follows: in
Section~\ref{sec:representation} we summarize the compact
representation (for more details, we refer the reader
to~\cite{WADSpaper}); in Section~\ref{sec:parallel} we describe the parallel algorithm of
\cite{WADSpaper} and discuss the details of its implementation; in Section~\ref{sec:experiments} we describe our
experiments for construction and query, and present their results;
finally, in Secion~\ref{sec:conclusions}, we present our conclusions
and future work.

\section{Representation}
\label{sec:representation}

\begin{figure*}[t]
\centering
\includegraphics[height=.25\textheight]{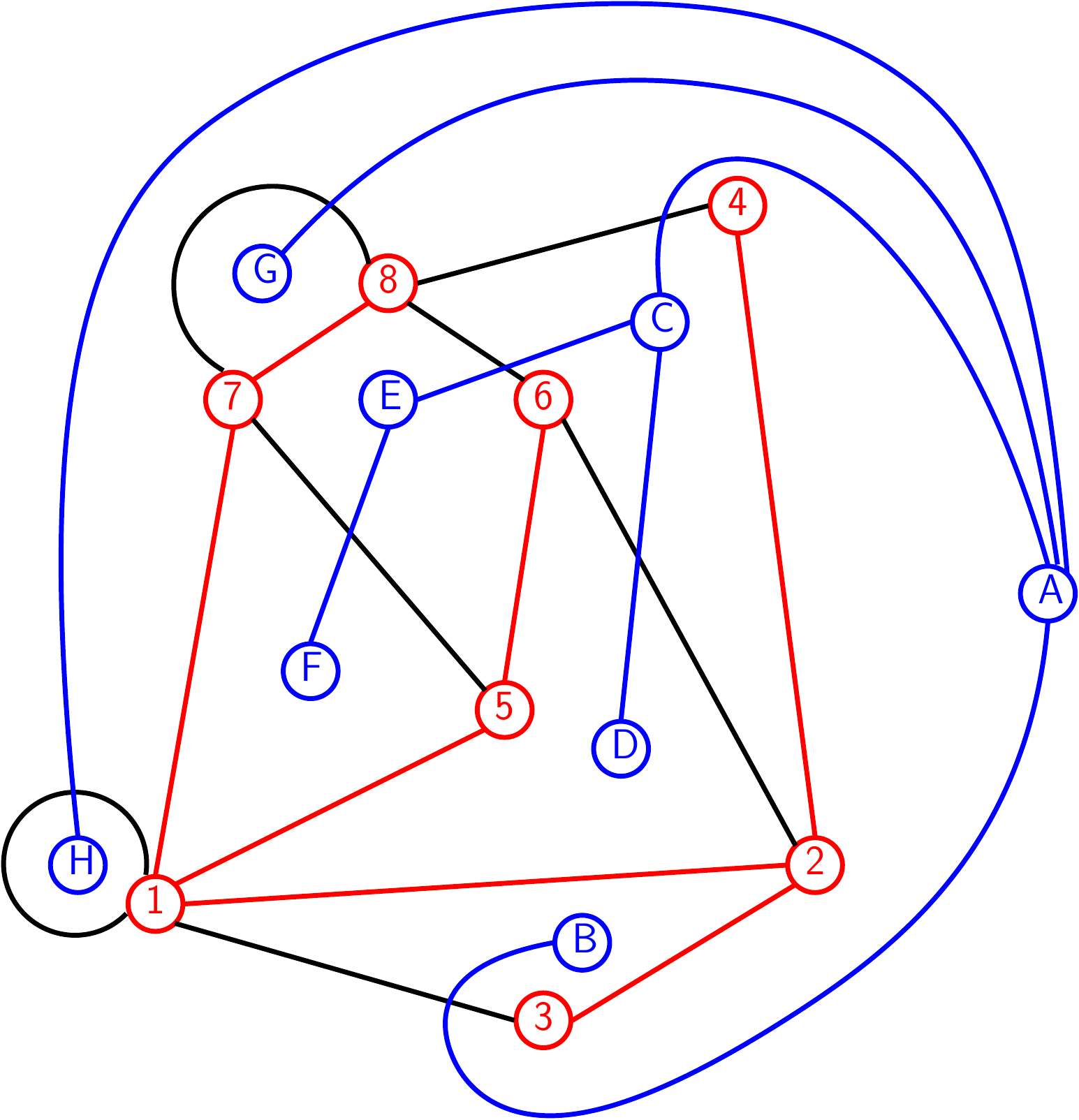}
\hspace{10ex}
\includegraphics[height=.25\textheight]{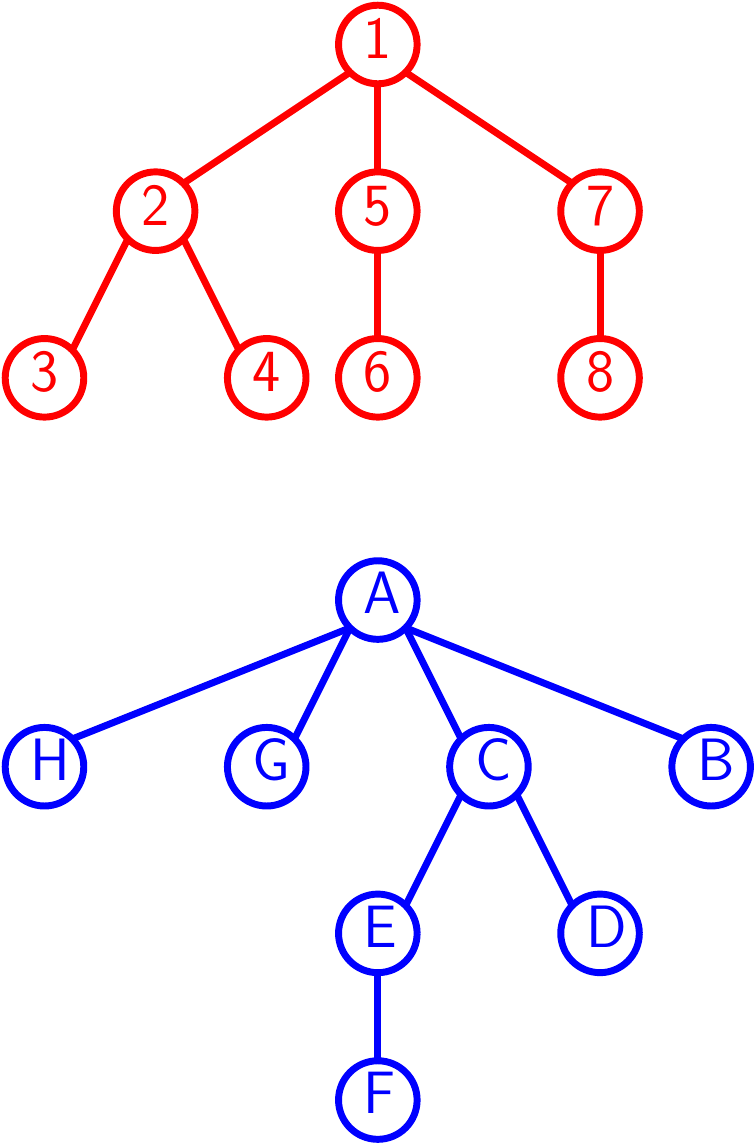}
\caption{{\bf Left:} A planar embedding of a planar graph $G$, with a spanning
  tree $T$ of $G$ shown in red and the complementary spanning tree $T^*$ of the
  dual of $G$ shown in blue.  {\bf Right:} The two spanning trees, with $T$
  rooted at the vertex {\sf 1} on the outer face and $T^*$ rooted at the vertex
  {\sf A} corresponding to the outer face.  We can represent this embedding of
  $G$ with the three bitvectors \(A [1..28] = 0110110101110010110100010100\),
  \(B [1..14] = 00101100110011\) and \(B^* [1..14] = 01001001110101\).}  
\label{fig:trees}
\end{figure*}

In~\cite{WADSpaper}, we introduced an extension of the compact
representation of Tur\'an~\cite{Turan1984} for connected planar
multi-graph with $n$ vertices and $m$ edges is introduced. The
extended representation works as follows: First, we chose an arbitrary
spanning tree $T$ of the planar graph, rooted at a vertex on the outer
face. Second, by traversing the spanning tree in DFS order, we create
three binary sequences $A$, $B$ and $B^*$. The first time we traverse
an edge of $T$, we write a $0$ in the sequence $B$, otherwise we write
a $1$ in $B$. Similarly, the first time that we reach an edge that
does not belong to $T$, we write a $0$ in $B^*$, otherwise we write a
$1$ in $B^*$. Finally, each time that we traverse an edge of $T$, we
write a $1$ in $A$, otherwise we write a $0$. The sequence $B$ has
length $2n-2$ and corresponds to the balanced-parentheses
representation of $T$. The sequence $B^*$ has length $2(m-n+1)$ and
corresponds to reversed balanced-parentheses representation of the
complementary spanning tree of the dual of the graph. The sequence $A$
has length $2m$ and indicates how the sequences $B$ and $B^*$ are
interleaved during the traversal. See Figure~\ref{fig:trees} as an
example of the construction of $A$, $B$ and $B^*$.

We can add sublinear number of bits to each
balanced-parentheses representation, we can support fast navigation in the
trees, and by storing the sequence $A$ as a bitvector, we can support fast
navigation in the graph. In particular, for the balanced-parentheses
representation we are interested on the constant time queries $\match$ and $\parent$, where
\(\match (i)\) returns the position of the parenthesis matching the $i$th
parenthesis and \(\parent (v)\) returns the parent of $v$, given as its
pre-order rank in the traversal, or 0 if $v$ is the root of its tree. For the
binary sequence $A$ we are interested in the constant time queries $\rank$ and
$\select$, where \(rank_b (i)\) 
returns the number of bits set to $b$ in the prefix of length $\ell$ of the $A$
and \(\select_b (j)\) returns the position of the $j$th bit set to $b$. For more
details on bitvectors and balanced-parentheses representations, we refer the
reader to Navarro's text~\cite{Navarro2016}. Based on the previous queries, we
defined the following basic queries:

\begin{description}
\item[\(\first (v)\):] return $i$ such that the first edge we process while
  visiting $v$ is the $i$th we process during our traversal; 
\item[\(\Next (i)\):] return $j$ such that if we are visiting $v$ when we
  process the $i$th edge during our traversal, then the next edge incident to
  $v$ in counterclockwise order is the one we process $j$th; 
\item[\(\mate (i)\):] return $j$ such that we process the same edge $i$th and
  $j$th during our traversal; 
\item[\(\vertex (i)\):] return the vertex $v$ such that we are visiting $v$ when
  we process the $i$th edge during our traversal. 
\end{description}
The compact representation support these basic queries in constant time as follows:
\begin{eqnarray*}
\first (v) & = & \left\{ \begin{array}{l@{\hspace{1.5ex}}l}
	A.\select_1 (B.\select_0 (v - 1)) + 1 & \mbox{if \(m \geq 1\)} \\
	0 & \mbox{otherwise}
\end{array} \right. \\[2ex]
\Next (i) & = & \left\{ \begin{array}{l@{\hspace{1.5ex}}l}
	i + 1 & \mbox{if \(A [i] = 0\) and \(i < 2 m\)} \\
	\mate (i) + 1 & \mbox{if \(A [i] = 1\) and \(B [A.\rank_1 (i)] = 0\)} \\
	0 & \mbox{otherwise}
\end{array} \right. \\[2ex]
\end{eqnarray*}
\begin{eqnarray*}
\mate (i) & = & \left\{ \begin{array}{l@{\hspace{1.5ex}}l}
	A.\select_0 (B^*.\match (A.\rank_0 (i))) & \mbox{if \(A [i] = 0\)} \\
	A.\select_1 (B.\match (A.\rank_1 (i))) & \mbox{otherwise}
\end{array} \right. \\[2ex]
\vertex (i) & = & \left\{ \begin{array}{l}
	B.\rank_0 (A.\rank_1 (i)) + 1\\\hspace{3ex} \mbox{if \(A [i] = 0\) and \(B [A.\rank_1 (i)] = 0\)} \\[1ex]
	B.\parent (B.\rank_0 (B.\match (A.\rank_1 (i)))) + 1\\\hspace{3ex} \mbox{if \(A [i] = 0\) and \(B [A.\rank_1 (i)] = 1\)} \\[1ex]
	B.\parent (B.\rank_0 (A.\rank_1 (i))) + 1\\\hspace{3ex} \mbox{if \(A [i] = 1\) and \(B [A.\rank_1 (i)] = 0\)} \\[1ex]
	B.\rank_0 (B.\match(A.\rank_1 (i))) + 1\\\hspace{3ex} \mbox{otherwise.}
\end{array} \right.
\end{eqnarray*}

With those basic queries, it is possible to define more complex queries. As an example,
we present three queries based on the basic ones: $\counting(v)$, the number of
neighbors of vertex $v$; $\listing(v)$, the list of neighbors of vertex $v$, in
counterclockwise order; $\face(e)$, the list of vertices, in clockwise
order, of the face where the edge $e$ belongs. The implementation of the three
queries follows. For queries $\listing(v)$ and $\face(e)$, we report each vertex
by using the function {\tt print}.

\begin{figure}[h]
  \centering
  \begin{minipage}[t]{0.32\textwidth}
  \begin{function}[H]
%%    \footnotesize
    \small
    \DontPrintSemicolon
    \SetVlineSkip{0.5ex}
    \LinesNumbered
    \SetKwFunction{Next}{next}
    \SetKwFunction{First}{first}
    \SetKwInOut{Input}{Input}
    % I/o
    \Input{node $v$}
    \BlankLine
    $d \asgn 0$\;
    $nxt \asgn \First(v)$\;
    \While{$nxt < 2m$} {
      $nxt \asgn \Next(nxt)$\;
      $d \asgn d + 1$\;
    }
    \caption{counting()}
    \label{func:counting}
  \end{function}
\end{minipage}%
\hspace{.4em}%
\begin{minipage}[t]{0.32\textwidth}
  \begin{function}[H]
    \DontPrintSemicolon
    \SetVlineSkip{0.5ex}
    \small
    \LinesNumbered
    \SetKwFunction{Next}{next}
    \SetKwFunction{First}{first}
    \SetKwFunction{Mate}{mate}
    \SetKwFunction{Vertex}{vertex}
    \SetKwFunction{Print}{print}
    %\footnotesize
    \SetKwInOut{Input}{Input}
    % I/o
    \Input{node $v$}
    \BlankLine
    $nxt \asgn \First(v)$\;
    \While{$nxt < 2m$} {
      $mt \asgn \Mate(nxt)$\;
      $\Print(\Vertex(mt))$\;
      $nxt \asgn \Next(nxt)$\;
    }
    \caption{listing()}
    \label{func:listing}
  \end{function}
\end{minipage}
\hspace{.4em}%
\begin{minipage}[t]{0.32\textwidth}
  \begin{function}[H]
    \DontPrintSemicolon
    \SetVlineSkip{0.5ex}
    \LinesNumbered
    \small
    \SetKwFunction{Next}{next}
    \SetKwFunction{First}{first}
    \SetKwFunction{Mate}{mate}
    \SetKwFunction{Vertex}{vertex}
    \SetKwFunction{Print}{print}
    \SetKwInOut{Input}{Input}
    % I/o
    \Input{edge $e$}
    \BlankLine
    $nxt \asgn e$,
    $flag \asgn true$\;
    \While{$nxt \neq e~or~flag$} {
      $flag \asgn false$\;
      $mt \asgn \Mate(nxt)$\;
      $\Print(\Vertex(mt))$\;
      $nxt \asgn \Next(nxt)$\;
    }
    \caption{face()}
    \label{func:face}
  \end{function}
\end{minipage}
\end{figure}

In the rest of this paper, we explain how to construct $A$,
$B$ and $B^*$ efficiently in parallel, and demostrate experimentally that our
representation is practical.

\section{Parallel algorithm for compact planar embeddings}
\label{sec:parallel}

In this section we discuss the parallel construction of the compact
representation of planar embeddings. Since the compact representation is based on spanning trees and tree
traversals, we can borrow ideas of well-known parallel algorithms, such as Euler
Tour traversal in parallel or parallel computation of spanning trees.

For the analysis of our algorithm, we use the {\em Dynamic Multithreading
  (DyM) Model} \cite{Cormen2009}. In this model, a multithreaded computation is
modelled as a directed acyclic graph (dag) where vertices are instructions and
an edge $(u,v)$ represents 
precedence among instruction $u$ and $v$. The model is based on two parameters
of the multithreaded computation: its {\em work} $T_1$ and its {\em
  span} $T_\infty$. The work is the running time on a single thread, that is,
the number of nodes (i.e., instructions) in the dag, assuming each instruction
takes constant time. The span is the length of the longest path in the dag; the
intrinsically sequential part of the computation. The time $T_p$ needed to
execute the computation on $p$ threads has complexity 
$\Theta(T_1/p+T_\infty))$, which can be reached with a greedy scheduler. The improvement of a multithreaded
computation using $p$ threads is called {\em speedup}, $T_1/T_p$. The upper
bound on the achievable speedup, $T_1/T_\infty$, is called {\em parallelism}.
Finally, the {\em efficiency} is defined as $T_1 /pT_p$ and can be interpreted as
the percentage of improvenment achieved by using $p$ cores or how close we
are to the linear speedup. In the DyM
model, the workload of the threads is balanced by using the {\em work-stealing}
algorithm \cite{Blumofe:1999:SMC:324133.324234}.

To describe parallel algorithms in the DyM model, we augment sequential
pseudocode with three keywords. The {\bf spawn} keyword, followed by a procedure
call, indicates that the procedure should run in its own thread and 
may thus be executed in parallel to the thread that spawned it. The {\bf sync}
keyword indicates that the current thread must wait for the termination of all
threads it has spawned. Finally, {\bf parfor} is ``syntactic sugar'' for {\bf
  spawn}ing one thread per iteration in a for loop, thereby allowing these
iterations to run in parallel, followed by a {\bf sync} operation that waits for
all iterations to complete. In practice, the {\bf parfor} keyword is implemented
by halving the range of loop iterations, spawning one half and using the current
procedure to process the other half recursively until reaching one iteration per
range. After that, the iterations are executed in parallel. Therefore, this
implementation adds an overhead bounded above the logarithm of the number of
loop iterations. In our algorithm, we include such overhead in our
complexities.

For the rest of this section, each tree $T$ is represented with an adjacency
list representation. Such representation consists of an array of nodes of size
$n$, $V_{T}$, and an array of edges of size $m$, $E_{T}$. Each node $v\in V_{T}$
stores two indices in $E_{T}$, $v.first$ and $v.last$, indicating the adjacency
list of $v$, sorted counterclockwise around $v$ and starting with $v$'s parent
edge (except the root). Notice that the number of children of $v$ is
$(v.last-v.first)$. Each edge $e\in E_{T}$ has three fields, $e.src$, which is a
pointer to the source vertex, $e.tgt$, which is a pointer to the target vertex
and $e.cmp$, which is the position in $E_{T}$ of the complement edge of $e$,
$e^{'}$, where the $e^{'}.src = e.tgt$ and $e^{'}.tgt = e.src$. For $x\in
\{e.src, e.tgt\}$, we use $\mathit{next}(x)$ and $\mathit{first}(x)$ to denote
the indices in $E_{G}$ of $e$'s successor and of the first element (parent edge)
in $x$'s adjacency list, respectively. Both are easily computed in constant time
by following pointers. Notice that $|E_{T}|=2(|V_{T}|-1)$. The representation of
graphs is similar, with the exception that the concept of \emph{parent} of a
vertex is not valid in graphs, therefore the first edge in the adjacency list of
a vertex $v$ cannot be interpreted as the $v$'s parent edge.

\subsection{Parallel construction of compact planar embeddings}
\label{subsuc:paralgo}
Given a planar embedding of a connected planar graph
$G=(V_G,E_G)$, for the moment we assume a spanning tree of $T=(V_T,E_T)$ of $G$ and an
array $C$ that stores the number of edges of $G\setminus T$ between two
consecutive edges in $T$, in counterclockwise order, are given as part of input. Later,
in Section \ref{subsec:spanning}, we explain how to obtain the spanning tree $T$
and the array $C$ in parallel. With the spanning tree, we construct the bitvectors $A$, $B$ and $B^*$ by
performing an Euler Tour over $T$. During tour, by
writing a $0$ for each forward (parent to child) edge and a $1$ for each
backward (child to parent) edge, we obtain the bitvector $B$, by counting the
number of edges of $G\setminus T$ between two consecutive edges of $T$ (array
$C$), representing them by $0$'s and the edges of $T$ by $1$'s, we obtain the
bitvector $A$, and by using the previous Euler Tour and the array $C$ we can
obtain the bitvector $B^*$. Algorithm \ref{algo:parAlgo} shows this idea in more
details. The algorithm works in six steps: In the first step, the algorithm
creates an auxiliar array $LE$ (line 4) that is used to store the traversal of
the tree following the Euler Tour. Each entry of $LE$ represents one traversed
edge of $T$ and stores four fields: {\em value} is $0$ or $1$ depending on
whether the edge is a forward or a backward edge, respectively; {\em succ} is
the index in $LE$ of the next edge in the Euler tour; {\em rankA} is the rank of
the edge in $A$; and {\em rankB} is the rank of the edge in $B$.

\begin{algorithm2e}
  \footnotesize
  % keywords
  \SetKwInOut{Input}{Input}
  \SetKwInOut{Output}{Output}
  \SetKwFor{PFor}{parfor}{do}{end}
  \SetKwFunction{parListRanking}{parallelListRanking}
  \SetKwFunction{createRS}{createRankSelect}
  \SetKwFunction{createBP}{createBP}
  \LinesNumbered
  \DontPrintSemicolon
  \SetVlineSkip{0.5ex}
  \SetCommentSty{textit}
  % I/o
  \Input{A planar embedding of a planar graph $G=(V_{G},E_{G})$, a spanning
    tree $T=(V_{T},E_{T})$ of $G$, an array $C$ of
    size $|E_T|$, the starting vertex $\mathit{init}$ and the number of threads,
    $\mathit{threads}$.} 
  \Output{Bitvectors $A$, $B$ and $B^*$ induced by $G$ and $T$.}
  \BlankLine
  % algorithm
  $\mathit{A} \asgn {}$a bitvector of length $|E_{G}|$\;
  $\mathit{B} \asgn {}$a bitvector of length $|E_{T}|-2$\;
  $\mathit{B^*} \asgn {}$a bitvector of length $|E_{G}|-|E_{T}|+2$\;
  $\mathit{LE} \asgn {}$an array of length $|E_{T}|$\;
  $\mathit{chk} \asgn |E_{T}|/\mathit{threads}$\;

  \PFor{$t \asgn 0$ \KwTo $\mathit{threads}-1$}{
    \For{$i \asgn 0$ \KwTo $\mathit{chk}-1$}{
      $j \asgn t*\mathit{chk} + i$\;

      $\mathit{LE}[j].\mathit{rankA} \asgn C[E_{T}[j].cmp]+1$\;
      $\mathit{LE}[j].\mathit{rankB} \asgn 1$\;
      
      \eIf(\tcp*[h]{forward edge}){$E_{T}[j].src==\mathit{init}$ OR
        $\mathit{first}(E_{T}[j].src)\neq j$}{
            $\mathit{LE}[j].\mathit{value} \asgn 0$ \tcp*[h]{opening parenthesis}

            \eIf{$E_{T}[j].tgt$ is a leaf} {
                   $\mathit{LE}[j].\mathit{succ} \asgn E_{T}[j].cmp$\;
            }
            {
                   $\mathit{LE}[j].\mathit{succ} \asgn \mathit{first}(E_{T}[j].tgt)+1$\;
            }
      }(\tcp*[h]{backward edge})
      {
            $\mathit{LE}[j].\mathit{value} \asgn 1$ \tcp*[h]{closing parenthesis}

            \eIf{$E_{T}[j]$ is the last edge in the adjacency list of
                 $E_{T}[j].src$}{ 
                 $\mathit{LE}[j].\mathit{succ} \asgn \mathit{first}(E_{T}[j].tgt)$\;
            }
            {
                $\mathit{LE}[j].\mathit{succ} \asgn \mathit{next}(E_{T}[j].tgt)$\;
            }
       }
    }
  }
  $\parListRanking(\mathit{LE})$\;
  
  \PFor{$t \asgn 0$ \KwTo $\mathit{threads} - 1$} {
    \For{$i \asgn 0$ \KwTo $\mathit{chk}-1$}{
      $j \asgn t*\mathit{chk}+i$\;
      $A[\mathit{LE}[j].\mathit{rankA}] \asgn 1$ \tcp*[h]{By default, all elements of
        $\mathit{A}$ are 0's}\;
      $B[\mathit{LE}[j].\mathit{rankB}] \asgn \mathit{LE}[j].\mathit{value}$\;
    }
  }

  $\mathit{D}_{pos}, \mathit{D}_{edge} \asgn {}$two arrays of length
  $|E_{G}|-|E_{T}|+2$\;
  \PFor{$t \asgn 0$ \KwTo $\mathit{threads} - 1$} {
    \For{$i \asgn 0$ \KwTo $\mathit{chk}-1$} {
      $j \asgn t*\mathit{chk}+i$\;
      $pos \asgn \mathit{LE}[j].\mathit{rankA}-\mathit{LE}[j].\mathit{rankB}$\;
      $lim \asgn \mathit{ref}(E_{T}[j].cmp)+C[E_{T}[j].cmp]$\;
      \For{$k \asgn \mathit{ref}(E_{T}[j].cmp)+1$ \KwTo $lim$} {
        $\mathit{D}_{pos}[k] \asgn pos$\;
        $\mathit{D}_{edge}[pos] \asgn k$\;
        $\mathit{pos} \asgn \mathit{pos}+1$
      }
    }
  }
  $\mathit{chk} \asgn (|E_{G}|-|E_{T}|+2)/\mathit{threads}$\;
  \PFor{$t \asgn 0$ \KwTo $\mathit{threads} - 1$} {
    \For{$i \asgn 0$ \KwTo $\mathit{chk}-1$}{
      $j \asgn t*\mathit{chk}+i$\;
      $cmp \asgn E_{T}[\mathit{D}_{edge}[j]].cmp$\;
      \If{$j > \mathit{D}_{pos}[cmp]$} {
        $B^{*}[j+1] \asgn 1$ \tcp*[h]{By default, all elements of
        $\mathit{B^*}$ are 0's}\;
      }
    }
  }
  $\createRS(\mathit{A})$, $\createBP(\mathit{B})$, $\createBP(\mathit{B^*})$\;
  \caption{Parallel compact planar embedding algorithm (\parAlgo).}
  \label{algo:parAlgo}
\end{algorithm2e}

In the second step, the algorithm traverses $T$
(lines 6--22). For each edge $e_j\in E_T$, {\em rankA} is set as
$C[E_{T}[j].cmp]+1$ and {\em rankB} as $1$ (lines 9--10). Those ranks will be
used later to compute the final position of the edges in $A$, $B$ and $B^*$. For
each forward edge, a $0$ is written in the corresponding {\em value} field and
the {\em succ} field is connected to the next edge in the Euler Tour. For
backward edges is similar. Considering the adjacency list representation of $T$,
all the edges in the adjacency list of a node (except the root) of $T$ are
forward edges, except the first one (parent edge). For the root, all
the edges of its adjacency list are forward edges.

In the third step, the algorithm computes the final ranks in $A$ and $B$ using a
parallel list ranking algorithm (line 23). We use the algorithm introduced in
\cite{Helman2001265} over the {\em rankA} and {\em rankB} fields of $LE$ to
obtain the final position of each edge in $A$ and $B$, respectively.

In the fourth step, bitvectors $A$ and $B$ are written. If initially all the
elements of $A$ are $0$'s, it is enough to set to $1$ all the elements given by
the fields {\em rankA}'s. For $B$, the algorithm copies the content of field
{\em value} at position {\em rankB}, for all the elements in $LE$.

In the fifth step, the algorithm computes the position of each edge of
$G\setminus T$ in $B^*$. That information is implicit in the fields {\em rankA}
and {\em rankB} of $LE$ (line 33), after the list ranking of the third step. For
each edge $e_j\in E_T$, the algorithm 
computes the positions, in $B^*$, of the edges in $G\setminus T$ that follow, in
counterclockwise order, the complement edge of $e_j$ (lines 34--38). The
algorithm uses two auxiliar arrays, $\mathit{D}_{pos}$ and $\mathit{D}_{edge}$. The
entry $\mathit{D}_{pos}[j]$ stores the position of the edge $e_j$ of $G\setminus
T$ in $B^*$. The array $\mathit{D}_{edge}$ is the inverse of
$\mathit{D}_{pos}$. It stores the position of the $j$-th edge of $B^*$ in 
$G\setminus T$. Thus, $\mathit{D}_{pos}[i]=j$ iff $\mathit{D}_{edge}[j]=i$. In
this step, the function $\mathit{ref}(E_{T}[j])$ returns the position of the
edge $e_j$ of $E_T$ in $E_G$.

In the sixth step, the algorithm computes if the edges stores in
$\mathit{D}_{pos}$ are forward or backward edges. For each edge $e$ in
$G\setminus T$, it is done by comparing the position in $B^*$ of $e$ and
its complement. If the position of $e$ is greater than the position of its
complement, then $e$ is a backward edge and, therefore, represented by a $1$. By
default, we assume that all the elements of $B^*$ are $0$'s.

Finally, the structures to support operations $\rank$, $\select$, $\match$ and
$\parent$ are constructed. For the bitvector $A$, Algorithm \ref{algo:parAlgo}
uses the parallel algorithms of Labeit et al. \cite{7786147}
($\mathtt{createRankSelect}$). In the case of $B$ and $B^*$, the algorithm uses
the parallel algorithm of Ferres et al. \cite{Ferres2015} ($\mathtt{createBP}$)
for balanced parenthesis sequences.

Now we present the analysis of our algorithm. In the first step there is not
computation involved, therefore, we do not include it in the complexity. In the
second step, the algorithm 
traverse the edges of $T$, performing an independent computation in each edge,
therefore, with the overhead of the {\bf parfor} loop, we obtain $T_1=O(n)$ and
$T_\infty=O(\lg n)$ time. The third step uses the
algorithm of Helman and J\'aJ\'a for the parallel list ranking problem over $m$ elements, with
complexities $T_1=O(n)$ and $T_\infty=O(\lg n)$ time. In the fourth step, the
assignation of the values to $A$ and $B$ can be done independently for each
entry of the bitvectors. With the overhead of the parallel loop, we have
$T_1=O(n)$ and $T_\infty=O(\lg n)$ time. In the 
fifth step, the algorithm traverses all the edges in $G\setminus T$. Observe
that the range of the loop in line 35 can be processed in parallel
using a domain decomposition technique. With that, we obtain $T_1=O(m-n)$ and
$T_\infty=O(\lg(m-n))$ time. We decide to implement it as it appears in line 35, because we
obtained good practical results. If we need to process the line 35 in parallel,
we can use the same domain decomposition technique of lines 30--32. Similar to
the fourth step, in the sixth step the algorithm sets the entries of the
bitvector $B^*$, which can be done independently for each entry. Therefore,
$T_1=O(m-n)$ and $T_\infty=O(\lg(m-n))$ time. 

The rank/select structures can be constructed in $T_1=O(m)$ and
$T_\infty=O(\lg m)$ time by using the results of \cite{7786147}. A structure
that supports $\match$ and $\parent$ operations over a balanced parentheses
sequence can be constructed in $T_1=O(m)$ and $T_\infty=O(\lg m)$ time with the
results of \cite{Ferres2015}.

In addition to the size of the compact data structure, the memory consumption
of our algorithm depends on the size of arrays $LE$, $D_{pos}$ and
$D_{edge}$. The array $LE$ uses $O(n\lg n)$ bits, and arrays $D_{pos}$ and
$D_{edge}$ uses $O((m-n)\lg n)$ bits. Thus, the total memory consumption of our
algorithm is $O(m\lg n)$ bits plus the output data structure. Notice that the
memory comsumption is independent of the number of threads.

In summary, we have the following theorem.

\begin{theorem}
  \label{theo:theo1}
  Given a planar embedding of a connected planar graph $G=(V_G,E_G)$ with $m$
  edges and a spanning tree of $G$, we can compute in parallel compact
  representation of $G$, using $4m + o(m)$ bits and supporting navegational
  operations, in $O(m)$ work, $O(\lg m)$ span, $O(m/p+\lg m)$ time, using
  $O(m\lg n)$ bits of additional memory, where $p$ is the number of available threads.
\end{theorem}

\subsection{Parallel computation of spanning trees}
\label{subsec:spanning}

In this section we discuss the parallel computation of the spanning tree $T=(V_T,E_T)$ and
the array $C$ used in Section \ref{subsuc:paralgo}.

The work of Bader and Cong \cite{BaderCong2005} can be used to compute a
spanning tree of a planar graph. Their algorithm works as follows:
Given a starting vertex of the graph $G$ with $n$ vertices and $m$ edges, the algorithm computes
sequentially a spanning tree of size $O(p)$, called {\em stub spanning
  tree}, where $p$ is the number of available threads. Then, an evenly number of
leaves of the stub spanning tree are assigned 
to the $p$ threads as starting vertices. Each thread traverses $G$, using its
starting vertices, constructing spanning trees with a DFS traversal using a
stack. For each vertex, a reference to its parent is assigned. Since a vertex can be visited by
several threads, the assigment of the parent of the vertex may genarate a {\em race
  condition}. However, since the parent assigned by any thread already
belongs to a spanning tree, any assignation will generate a correct tree. Thus,
the race condition is a {\em benign} race condition. Once a thread has no more
vertices on its stack, the thread tries to steal vertices from the stack of other
thread by using the work stealing algorithm. Since the spanning
trees generated by all the threads are connected to the stub spanning tree, the
union of all the spanning tree generates a spanning tree of $G$. Thus, the
algorithm gives an array of parent references for each vertex.
With such array of references, we can construct the corresponding adjacency list
representation of the spanning tree. To do that, we mark with a
$1$ each edge $E_G$ that belongs to $E_T$ and with $0$ the rest of the
edges. Using a parallel prefix sum algorithm over $E_G$, we compute the position
of all the marked edges of $E_G$ in $E_T$. The $\mathit{first}$ and
$\mathit{last}$ fields of each node in the spanning tree are computed
similarly. As a byproduct of the computation of $E_T$, we can compute an array
$C$ which stores the number of edges of $G\setminus T$ between two consecutive
edges in $T$, in counterclockwise order. It can be done by using the marks in
the edges, counting the number of $0$'s between two consecutive $1$'s. Notice
the starting vertex for the stub spanning tree must be in the outer face of $G$,
to meet the description of the compact data structure for planar embeddings.

The complexities of the spanning tree algorithm depends on both the random
traversal of the threads and the diameter of $G$: $T_1=O(m+n)$ work,
$T_\infty=O(m+n)$ span, since the stub spanning tree is computed sequentially
and its size is proportional to the number of threads, and $T_p=O((m+n)/p)$ 
expected time for general random graphs, for $p\ll n$. This algorithm has a worst case when $G$ has
diameter of $O(m)$ and low connectivity. In that case, the expected time is
$T_p=(m+n)$. Despite its span and worst case, the algorithm of Bader
and Cong has a good practical behavior and its implementation is simple. The
adjancency list representation of the spanning tree $T$ and the array $C$ are
computed by using a parallel prefix sum algorithm, which is well-known to have
$T_1=O(m)$ and $T_\infty=O(\lg m)$ time.

Since the spanning tree of $G$ is part of the input of our algorithm for planar
embeddings, we decide to explore different parallel models to overcome the worst
case of Bader and Cong's algorithm, and thus improve the overall complexity of
our algorithm. The algorithm introduced in this manuscript is simple
enough to be adapted to other parallel models, because it is based on Euler
Tour and list ranking algorithm, and parallel filling of arrays, which are well-known
algorithms in most of the parallel models. In particular, we can adapt our
parallel algorithm to the CRCW PRAM model, since we can use the CRCW PRAM Euler
Tour algorithm of \cite{1327954}, with $O(\lg n)$ parallel time using $O(m)$
cores, and the CRCW PRAM list ranking algorithm of \cite{COLE1989334}, with
$O(\lg n)$ parallel time using $O(n/\lg n)$ cores. For spanning trees, we
have the algorithms of \cite{1676869} and \cite{SHILOACH19825} in the CRCW PRAM
model. The algorithm in \cite{1676869} takes $O(\lg n)$ parallel time using
$O(m)$ cores, and the algorithm in \cite{SHILOACH19825} takes $O(\lg n)$
parallel time using $n+2m$ cores.

Thus, in the CRCW PRAM model, our algorithm for
compact planar embeddings reaches logarithmic parallel time, including the
computation of the spanning tree. The array $C$ can be constructed using the
strategy explained before, based on parallel prefix sum.

Despite its the worst case, we used the spanning tree algorithm of Bader and
Cong in our implementation and experiments, since it showed good practical
results. 

\section{Experiments}
\label{sec:experiments}

We implemented the \parAlgo~algorithm in C and compiled it using GCC 5.4 with
Cilk Plus extension, an implementation of the DyM model.\footnote{The code
and data needed to replicate our results are available
at \url{http://www.dcc.uchile.cl/~jfuentess/pemb/}.} In our implementation of
the parallel spanning tree algorithm of Bader and Cong, to reduce the worst
case, we included a treshold of $O(n/p)$ elements in the stack size of each
thread. Each time that a thread has more nodes that the threshold, that thread
create a new parallel task with the half of its stack. Additionally, we also
return for each node the reference to 
its parent. Returning the references to parents gives better performance than
forcing the first edge of each node to be the reference to its parent. However,
in the description of the \parAlgo~algorithm we assume that the first edge of
each node is the reference to its parent to make it more readable. Additionally,
we implemented a sequential algorithm called {\tt seq}, which corresponds to the
serialization of the \parAlgo~algorithm and a sequential DFS algorithm to
compute the spanning tree. To serialize a parallel algorithm in the DyM model,
we replaced each {\bf parfor} keyword for the {\bf for} keyword and deleted the
{\bf spawn} and {\bf sync} keywords.

\begin{figure}[t]
  \begin{minipage}[c]{0.5\textwidth}
   \centering
   \small
   \begin{tabular}{r@{\hspace{3ex}}l@{\hspace{3ex}}r@{\hspace{3ex}}r@{\hspace{3ex}}}
     \hline 	
     & Dataset & Vertices ($n$) & Edges ($m$) \\ 
     \hline 	
     1 & wc & 2,243,467  & 6,730,395 \\ 
     2 & pe5M & 5,000,000  & 14,999,983 \\ 
     3 & pe10M & 10,000,000  & 29,999,979 \\ 
     4 & pe15M & 15,000,000  & 44,999,983 \\ 
     5 & pe20M & 20,000,000  & 59,999,975 \\ 
     6 & pe25M & 25,000,000  & 74,999,979 \\ 
     \hline 	
   \end{tabular} 
   \vspace{1ex}
   \captionof{table}{Datasets used in the experiments of the
     \parAlgo~algorithm.\label{tbl:datasets}} 
    \vspace{1ex}
   \includegraphics[width=\linewidth]{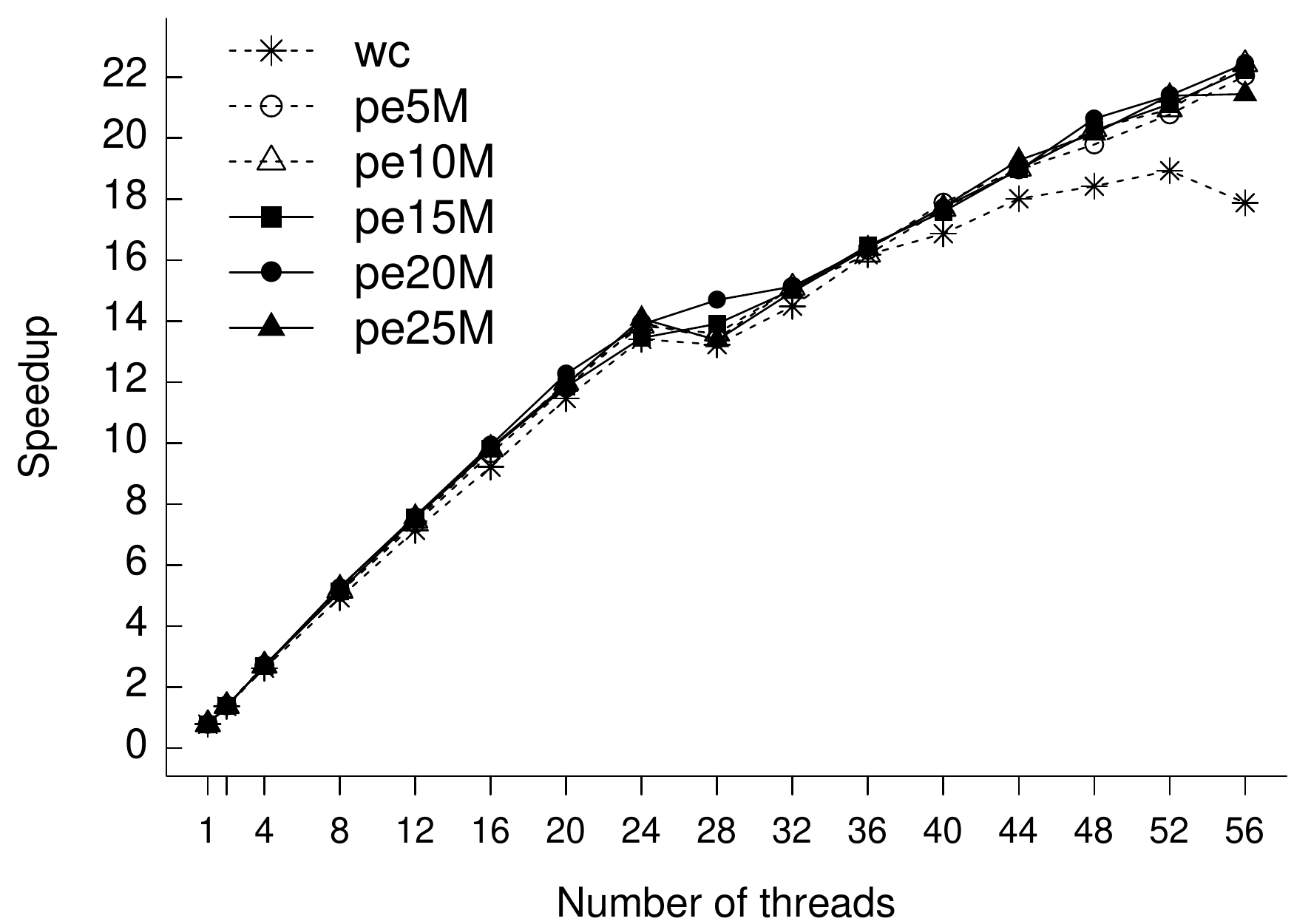}
  \end{minipage}
  \hspace{\stretch{1}}%
  \begin{minipage}[b]{0.47\textwidth}
    \setlength{\tabcolsep}{0pt}
    \begin{tabular}{c@{\hspace{.8em}}r@{ }r@{ }r@{ }r@{ }r@{ }r}
      \hline
      $p$ & {\tt wc} & {\tt pe5M} & {\tt pe10M} & {\tt pe15M} & {\tt pe20M}
      & {\tt pe25M}\\
      \hline
          {\tt seq} & 5.56 & 13.55 & 27.86 & 42.34 & 57.04 & 71.76 \\
          1 & 7.08 & 17.14 & 35.63 & 54.21 & 73.11 & 92.17 \\ 
          2 & 4.06 & 9.71 & 20.26 & 30.67 & 41.37 & 52.16 \\ 
          4 & 2.12 & 4.98 & 10.31 & 15.75 & 21.17 & 26.70 \\ 
          8 & 1.13 & 2.65 & 5.36 & 8.25 & 10.81 & 13.61 \\ 
          12 & .78 & 1.84 & 3.74 & 5.61 & 7.50 & 9.46 \\ 
          16 & .60 & 1.40 & 2.84 & 4.32 & 5.74 & 7.28 \\ 
          20 & .49 & 1.15 & 2.33 & 3.57 & 4.65 & 6.02 \\ 
          24 & .41 & .97 & 2.01 & 3.15 & 4.11 & 5.10 \\ 
          28 & .42 & 1.01 & 2.05 & 3.04 & 3.88 & 5.36 \\ 
          32 & .38 & .90 & 1.85 & 2.82 & 3.77 & 4.80 \\ 
          36 & .34 & .83 & 1.72 & 2.57 & 3.48 & 4.38 \\ 
          40 & .33 & .76 & 1.58 & 2.41 & 3.22 & 4.04 \\ 
          44 & .31 & .71 & 1.47 & 2.23 & 3.01 & 3.72 \\ 
          48 & .30 & .68 & 1.37 & 2.09 & 2.76 & 3.56 \\ 
          52 & .29 & .65 & 1.33 & 2.01 & 2.67 & 3.36 \\ 
          56 & .31 & .61 & 1.24 & 1.90 & 2.54 & 3.35 \\ 
      \hline
    \end{tabular}
  \end{minipage}%

  \leavevmode\begin{minipage}[t]{0.5\textwidth}
    \captionof{figure}{Speedup of the \parAlgo~algorithm.\label{fig:speedup}}
  \end{minipage}%
  \hspace{\stretch{1}}%
  \begin{minipage}[t]{0.47\textwidth}
    \captionof{table}{Running times of \parAlgo~algorithm in
      seconds.\label{tbl:parallelTimes}}
  \end{minipage}
\end{figure}

The experimental trials consisted in running the implementation on artificial
datasets of different number of nodes and threads. The datasets are shown in
Table \ref{tbl:datasets}. Each dataset was generated in three stages: In the
first stage, we used the function {\tt rnorm} of {\tt R} to generate random
coordinates $(x,y)$\footnote{The {\tt rnorm} function generates random numbers
  for the normal distribution given a mean and a standard deviation. In our
  case, the $x$ and $y$ components was generated using mean $0$ and standard
  deviation $10000$. For more information about the {\tt rnorm} function, please
  visit
  \url{https://stat.ethz.ch/R-manual/R-devel/library/stats/html/Normal.html}}. The
only exception was the dataset {\tt wc}, which corresponds to the
coordinates of $2,243,467$ uniques cities in the world.\footnote{The dataset
  containing the coordinates was created by MaxMind, available from
  \url{https://www.maxmind.com/en/free-world-cities-database}. The original
  dataset contains $3,173,959$ cities, but some of them have the same
  coordinates. We selected the $2,243,467$ cities with unique coordinates to
  build our dataset {\tt worldcities}.} In the second stage, we generated the
{\em Delaunay Triangulation} of the coordinates generated in the first
stage. The triangulations were generated using {\em Triangle}, a piece of
software dedicated to the generation of meshes and triangulations\footnote{The
  software is available at \url{
 http://www.cs.cmu.edu/~quake/triangle.html}. Our triangulations were generated  
  using the options {\tt -cezCBVPNE}.}. In the final stage, we generated planar
embeddings of the Delaunay triangulations computed in the second stage. The
planar embedding was generated with the {\em The Edge Addition Planarity
  Suite}\footnote{The suite is available at
  \url{https://github.com/graph-algorithms/edge-addition-planarity-suite}. Our
  embeddings were generated using the options {\tt -s -q -p}.}. The minimum and
maximum degree of the dataset {\tt wc} was 3 and 36, respectively. For the rest
of the datasets, the minimum degree was 3 and the maximum degree was 16. We
repeated each trial five times and recorded the median time.

The experiments were carried out on a NUMA machine with two NUMA nodes. Each
NUMA node includes a 14-core Intel\textregistered{} Xeon\textregistered{} CPU
(E5-2695) processor clocked at 
2.3GHz. The machine runs Linux 2.6.32-642.el6.x86\_64, in 64-bit mode. The
machine has per-core L1 and L2 
caches of sizes 64KB and 256KB, respectively and a per-processor
shared L3 cache of 35MB, with a 768GB DDR3 RAM memory (384GB per NUMA node),
clocked at 1867MHz. Hyperthreading was enabled, giving a total of 28 logical
cores per NUMA node.

Table \ref{tbl:parallelTimes} shows the running times obtained in our
experiments, and Figure \ref{fig:speedup} shows the speedups compared with the
{\tt seq} algorithm. On average, the {\tt seq} algorithm took about 76\% of the
time obtained by the \parAlgo~algorithm running with 1 thread. With $p\geq 2$, the
\parAlgo~algorithm shows better times than the {\tt seq} algorithm. We observe an
almost linear speedup up to $p=24$, with an efficiency of at least 56\%,
considering all the datasets. With $p=28$ the speedup has a slowdown, due to the
topology of our machine. Up to 24 cores, all the threads were running in
the same NUMA node. With $p\geq 28$, both NUMA nodes are used which implies higher
communication costs. The communication costs intra NUMA nodes are lower than the
communication costs inter NUMA nodes \cite{Drepper2007}. In particular, the case
of $p=28$ also uses both NUMA nodes, since at least one core on our machine was
available to OS processes. For $p=56$, the {\tt wc} dataset exhibits an
efficiency of only 32\% due to it is the smaller dataset. For the rest of
datasets, the lower efficiency was 38\%.

\begin{figure}[t]
\centering
 \begin{minipage}[c]{0.48\textwidth}
    \includegraphics[width=1.04\textwidth]{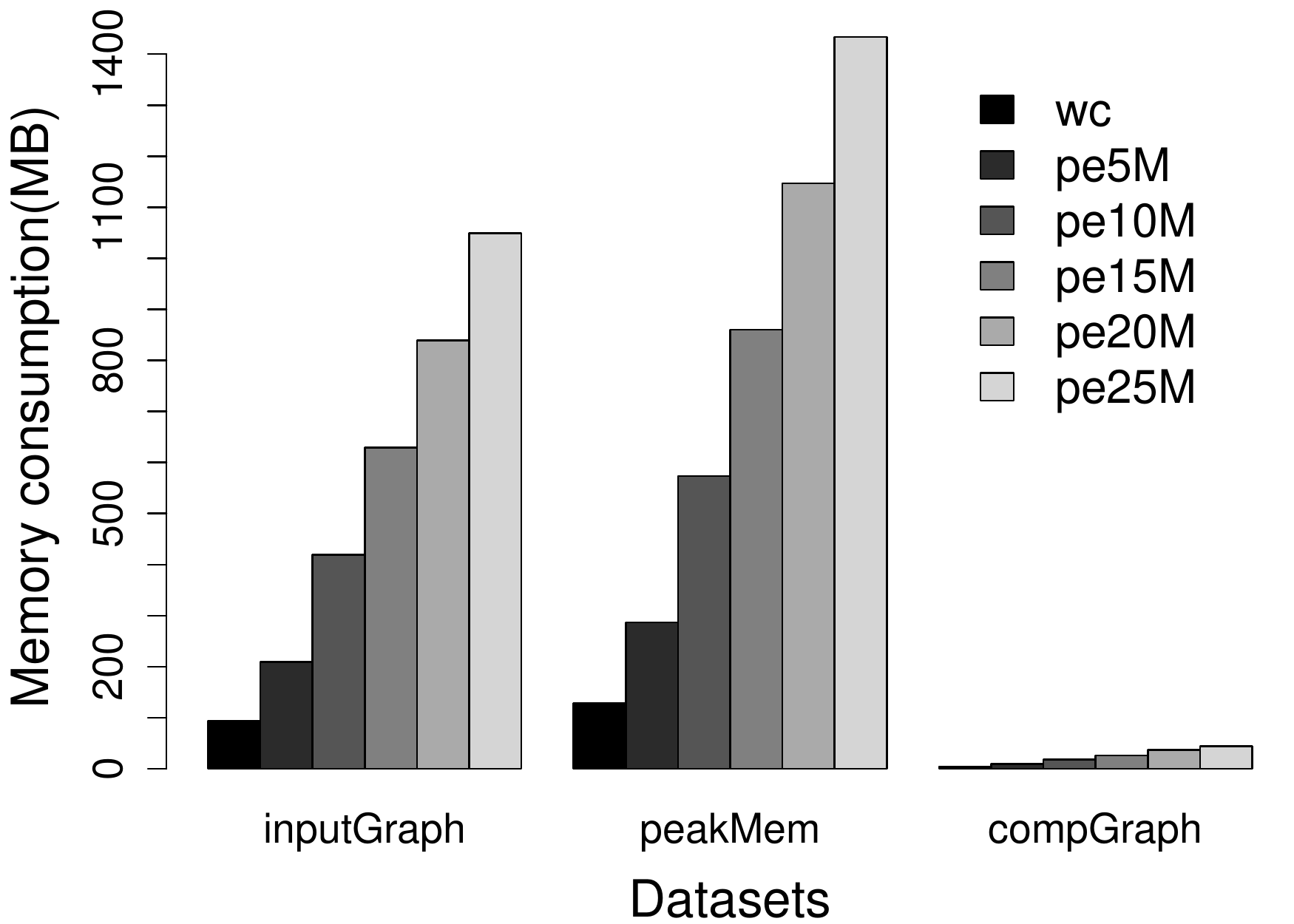}
 \end{minipage}%
 \hspace{\stretch{1}}%
 \begin{minipage}[c]{0.48\textwidth}
  \includegraphics[width=\linewidth]{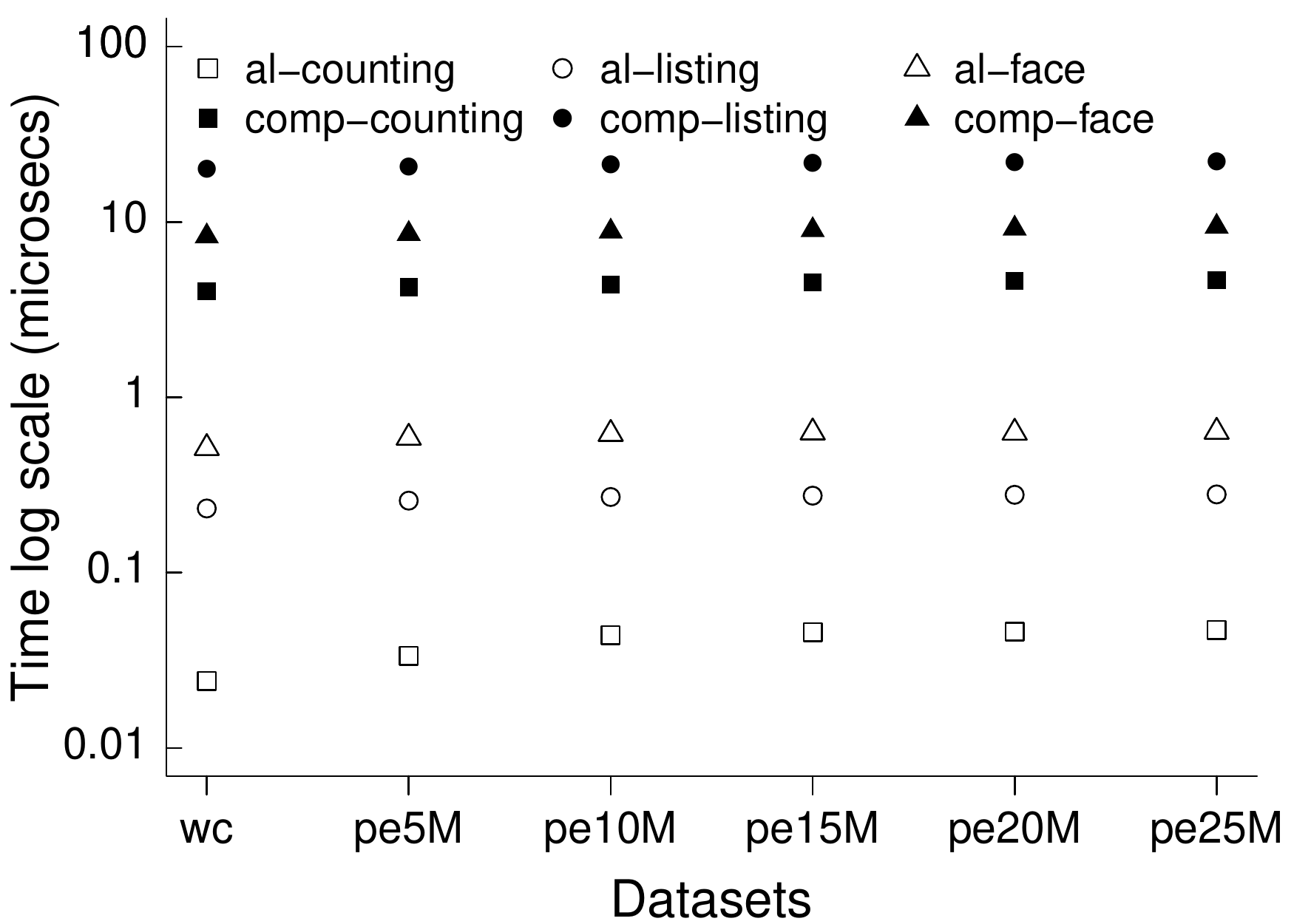}
 \end{minipage}\\[1ex]
 \leavevmode\begin{minipage}[t]{0.48\textwidth}
   \captionof{figure}{Memory consumption of the
     \parAlgo~algorithm. \label{fig:memory}}  
 \end{minipage}%
 \hspace{\stretch{1}}%
 \begin{minipage}[t]{0.48\textwidth}
    \captionof{figure}{Median times of counting, listing and face
      queries.\label{fig:queries-time}}
 \end{minipage}
\end{figure}

Figure \ref{fig:memory} shows the memory consumption of our
algorithm. Specifically, the figure shows the space used by each dataset in
adjacency list representation ({\tt inputGraph}), the peak of consumption of our
implementation ({\tt peakMem}) and the size of the compact representation of
each dataset ({\tt compGraph}). Compared with the space consumption of the
adjacency list representation, our implementation uses 36\% more space and the
compact representation uses about 4.6\% of it. The consumption per edge was
5 bits, which matches the Theorem \ref{theo:theo1}.

Finally, we tested the three queries introduced in Section
\ref{sec:representation}: {\tt counting}, {\tt listing} and {\tt face}. Observe
that, given the adjacency list 
representation described in Section \ref{sec:parallel}, to answer {\tt counting}
and {\tt listing} queries is straightforward. In our experiments, we tested {\tt counting}
and {\tt listing} 10 times for each vertex, and {\tt face} 10 times per edge. Figure \ref{fig:queries-time}
shows the median time per query, both for the adjacency list ({\tt
  al-counting}, {\tt al-listing} and  {\tt al-face}) and
compact representation ({\tt comp-counting}, {\tt comp-listing} and {\tt comp-face}). The
adjacency list representation allows to answer {\tt counting} and {\tt listing}
queries $100$ and $80$ times faster than the compact representation. This
result was expected, since the adjacency list representation we assumed already
has the list of neighbors in counterclockwise order. For the {\tt face} query,
the adjacency list representation is only $14$ times faster.

In summary, our parallel algorithm has good scalability to construct the compact
representation of planar embeddings of \cite{WADSpaper}. In particular, using
only one NUMA node, our algorithm scales linearly. To answer queries, compact
representations of data structures are slower than their
uncompacted counterparts. However, such compact representation use less memory,
allowing to fit data structures close to fast memories, such as main memory and
caches, speeduping up the overall computation for large datasets. In the
particular case of the {\tt face} query, a query closer to what we expect when
solving more realistic problems, our compact implementation is $14$ slower and uses
$20$ times less memory than the uncompacted representation.

\section{Conclusions}
\label{sec:conclusions}

In this paper, we presented the algorithm engineering of the parallel algorithm for the construction
of compact representations of planar embeddings introduced
in~\cite{WADSpaper}. We also
show empirically that our proposed implementation has good
scalability in shared-memory architectures. Finally, we tested three
different queries supported by our implementation and show that they
have good execution-time behavior, making them of practical
importance.

Notice, interestingly, that the compact representation can be extended
to {\em unconnected} planar graphs. To do this, we first need to find
all the connected components of the graph. Then, we compute an
arbitrary spanning tree for each connected component. Then, we
construct the binary sequences: the sequence $B$ will represent the
forest of the spanning trees, concatenating all the
balanced-parentheses representations; the sequence $B^*$ will
represent complementary spanning tree of the dual of the graph. Since
the outer face of all the connected components is the same, $B^*$
represents a tree and not a forest. Finally, sequence $A$ indicates
the interleaving of the sequences $B$ and $B^*$. For that, we
arbitrarily choose a connected component to start the traversal. For
the computation of the connected components, we can use the work of
Shun et al. \cite{Shun:2014:SPL:2612669.2612692} which has good
theoretical and practical results.

As future work, we will compare our implementation of the Bader and
Cong algorithm with the algorithm for connected components of Shun et
al. \cite{Shun:2014:SPL:2612669.2612692}. The algorithm of Shun et
al. computes the connected components of a graph by calling the
partitioning algorithm of Miller et
al.~\cite{Miller:2013:PGD:2486159.2486180}. During the partitioning
step, the algorithm performs multiple BFS's over the graph. All the
vertices in a partition are contracted to generate a new graph with
less edges. The process is recursively repeated until the contracted
graph has not edges. We can modify this algorithm in order to obtain
the spanning tree by returning the edges traversed in each BFS.

%%
%% Bibliography
%%

%% Either use bibtex (recommended), 

\bibliography{ESA}

%% .. or use the thebibliography environment explicitely

\end{document}